\documentclass[submission,copyright,creativecommons]{eptcs}
\providecommand{\event}{VPT 2019} %
\usepackage{breakurl}             %
\usepackage{underscore}        

\usepackage[utf8]{inputenc}
\usepackage{graphicx}
\usepackage{galois}

\usepackage{amsmath} 
\usepackage{algorithm}
\usepackage[noend]{algpseudocode}

\usepackage{alltt}
\usepackage{tikz}
\usepackage{etex}
\usepackage[etex=true,export]{adjustbox}
\usepackage{framed} %

\usepackage{amssymb}
\usepackage{upquote}
\usepackage{fancyvrb} %

\usepackage{float}

\usepackage{listings} 
	   \usepackage{courier}
\usepackage{paralist}
\usepackage{hyperref}

\newcommand{\true}{\mathsf{true}}
\newcommand{\false}{\mathsf{false}}

\newcommand{\C}{{\cal C}}

\newcommand{\abst}{\alpha}
\newcommand{\bodyfacts}{\mathsf{collect}}

\newcommand{\renames}{\rho}

\newcommand{\rename}{\mathsf{unfoldfold}}

\newcommand{\SAT}{\mathsf{SAT}}

\newcommand{\unfold}{{\sf pe}}

\newcommand{\Def}{\mathsf{Def}}

\newcommand{\anydim}[1]{\ge\! 0}

     {\end{tabular}\end{tt}\end{small}}
\def\anno#1{{\ooalign{\hfil\raise.07ex\hbox{\small{\rm #1}}\hfil%
        \crcr\mathhexbox20D}}}

\newtheorem{definition}{Definition}
\newtheorem{example}{Example}
\newtheorem{lemma}{Lemma}
\newtheorem{proposition}{Proposition}

\newcommand{\tuplevar}[1]{\mathbf{#1}}
 
\title{Polyvariant Program Specialisation \\with Property-based Abstraction }
\author{John P. Gallagher
\institute{Roskilde University, Denmark and
IMDEA Software Institute, Madrid, Spain}
\email{jpg@ruc.dk}
}
\def\titlerunning{Polyvariant Program Specialisation with Property-based Abstraction}
\def\authorrunning{J. P. Gallagher}

\begin{document}
\maketitle

\pagestyle{plain}
\pagestyle{myheadings}

\begin{abstract}

In this paper we show that property-based abstraction, an established
technique originating in software model checking, is a flexible method of
controlling polyvariance in program specialisation in a standard online
specialisation algorithm. Specialisation is a program transformation that
transforms a program with respect to given constraints that restrict its
behaviour. Polyvariant specialisation refers to the generation of two or
more specialised versions of the same program code. The same program point
can be reached more than once during a computation, with different
constraints applying in each case, and polyvariant specialisation allows
different specialisations to be realised. A property-based abstraction uses
a finite set of properties to define a finite set of abstract
versions of predicates, ensuring that only a finite number of specialised versions is
generated. The particular choice of properties is critical for
polyvariance; too few versions can result in insufficient specialisation,
while too many can result in an increase of code size with no corresponding
efficiency gains. Using examples, we show the flexibility of specialisation
with property-based abstraction and discuss its application in control flow
refinement, verification, termination analysis and dimension-based
specialisation.
\end{abstract}

\section{Program specialisation}
Specialisation is a program transformation that transforms a program with respect to some given constraints that restrict its behaviour. A classic example is the  loop in Figure \ref{fig-pe-exp}(a) for computing $\mathtt{z=x}^\mathtt{y}$. Figure  \ref{fig-pe-exp}(b) shows the result of specialising the loop with the input constraint $\mathtt{y=3}$, unfolding the loop three times and evaluating the statement $\mathtt{y}$\texttt{--} in the loop body.  
\begin{figure}[h]
\begin{center}
\begin{tabular}{l|l}
\begin{lstlisting}
z = 1;
while (y>0) {
  z = x*z;  y--;
}
\end{lstlisting} ~~~~~&~~~~~~~
\begin{lstlisting}
/* Input constraint y=3 */
z = 1; z = x*z; z = x*z;  z = x*z; 
\end{lstlisting}\\
~~~~~~~(a) & ~~~~~~~~~~~~~~(b)\\
\end{tabular}
\end{center}
\caption{Specialisation of a loop}
\label{fig-pe-exp}
\end{figure}

Some specialisation methods could further transform the code in Figure  \ref{fig-pe-exp}(b) to \texttt{z = x*x*x;}  using algebraic reasoning.
 As well as exploiting input constraints to specialise the program, we can perform internal specialisations based on 
constraints generated during program execution.  For example, when specialising a statement $\mathtt{if} (e) \{s_1\}\{s_2\}$, even where the test $e$ itself cannot be evaluated, the branch $s_1$ can be specialised with the constraint $e$ and the branch $s_2$ can be specialised with the constraint $\neg e$.  This is sometimes called \emph{driving} \cite{Turchin-88}.

 \emph{Polyvariant} specialisation refers to the generation of two or more specialised \emph{versions} of the same program code. 
For example suppose that the statement $\mathtt{if} (x<100) \{s_1\}\{s_2\}$ is  
reached twice during a computation, once with the constraint $x<100$ and the other with the constraint $x\ge 100$. Polyvariant specialisation gives rise to two instances of the statement in the specialised code, $s_1$ and $s_2$ respectively. 

Figure \ref{fig-polyvariance} illustrates both internal specialisation and polyvariance.
\begin{figure}[t]
\begin{center}
\begin{tabular}{l|l}
\begin{lstlisting}
while (x>0) {
   if (y<m) {
      y++;
   } else {
      x--;
   }
 }
\end{lstlisting} ~~~~~&~~~~~~~
\begin{lstlisting}
if (x>0) {
   while (y<m) {    /* x>0 */
      y++;}
   x--;
   while (x>0) {    /* y>=m */
      x--;}
}
\end{lstlisting}\\
&\\
~~~~~~~(a) & ~~~~~~~~~~~~~~(b)\\
\end{tabular}
\end{center}
\caption{Polyvariant specialisation of a loop}
\label{fig-polyvariance}
\end{figure}
The ``then" branch of the \textbf{if} statement in Figure \ref{fig-polyvariance}(a)
does not affect the loop condition and so if it is taken,  the ``then" branch is repeatedly taken until the test $\mathtt{y<m}$ fails.
Then the ``else" statement is executed (\texttt{x--;}) after which 
the ``else" branch is repeatedly taken, since it does not affect the condition $\mathtt{y<m}$, until the test $\mathtt{x>0}$ fails. 

Thus implicitly there are two distinct loops separated by \texttt{x--;} and this leads to the polyvariant specialisation shown in Figure \ref{fig-polyvariance}(b).  The loops in Figure \ref{fig-polyvariance}(b) are reconstructed from an internal control flow representation; the first \textbf{while} loop has a loop test corresponding to the \textbf{if} statement from the input program.  Further explanation of this example is given in Example \ref{pe-alg-ex}.

The main contribution of this paper is a specialisation algorithm that performs polyvariant specialisation. Instances of this algorithm have been previously used and briefly described \cite{DBLP:journals/tplp/KafleGGS18,DBLP:journals/tplp/KafleGG18,DomenechGG2018} but these papers did not present and discuss the general algorithm.
A key question is the control of polyvariance; in general there could be many (even an infinite number) of possible variants of a given
program point.  How does the specialisation algorithm determine a suitable set of variants, while ensuring termination of 
the specialisation algorithm.

The algorithm operates on constrained Horn clauses, which provide a representation language capable of representing the semantics of a wide range of programming languages and systems.  The algorithm is parameterised by a set of properties that control the generation of 
polyvariance.

\section{Preliminaries}\label{sec:prelim}
\subsection{Constraints and entailment}

Let $T$ be a theory and let $\C^T$ be the set of formulas (also called \emph{constraints}) constructed from the predicates and function symbols of $T$ together with variables and boolean connectives, and the formulas $\true$ and $\false$. $\models_T \phi$ means that $\phi$ is true in $T$, where  $\phi$ is a variable-free formula.  For example,
$\models_T 1 \ge 0$ where $T$ is the theory of linear real arithmetic (LRA).
In this paper, if we omit the theory subscript $T$ we assume that $T$ is the theory of linear real arithmetic, and we omit the symbol
$\models$ when clear from context.

Let $\phi \in \C^T$ be a constraint possibly containing variables; a substitution for the variables of $\phi$ is a \emph{grounding substitution} if the result of applying the substitution to $\phi$, say $\phi'$, 
contains no variables; if $\models_T \phi'$ the grounding substitution \emph{satisfies} $\phi$.

For all constraints $\phi$ and $\psi$, we say that $\phi$ \emph{entails} $\psi$ in $T$,  
written $\phi \preceq_T \psi$, if every grounding substitution that satisfies $\phi$ in $T$ also satisfies $\psi$ in $T$.   
For example, $x \ge 1 \preceq_T x \ge 0$ where $T$ is LRA.

We assume that there is a procedure called $\SAT_T$ such that for every $\phi \in \C^T$, $\SAT_T(\phi)$ is true if there is some substitution that satisfies $\phi$ and false otherwise.  
Using $\SAT_T$, we can check entailment;  $\phi \preceq_T \psi$ if and only if $\SAT(\phi \wedge \neg\psi)$ is false.

\begin{definition}[Generalisation]\label{def-gen}
A function $\renames: \C^T \rightarrow \C^T$ is called a \emph{generalisation operator} if $\phi \preceq_T \renames(\phi)$.
\end{definition}

\subsection{Constrained Horn clause representation of programs} 

A constrained Horn clause (CHC) over some constraint theory $T$ is a first-order predicate logic formula of the form $\forall \tuplevar{x_0} \ldots \tuplevar{x_k} 
(p_1(\tuplevar{x_1}) \wedge \ldots \wedge p_k(\tuplevar{x_k}) \wedge \phi \rightarrow
p_0(\tuplevar{x_0}))$, where $\phi$ is a finite conjunction of \emph{constraints} from $\C^T$, $\tuplevar{x_0},\ldots, \tuplevar{x_k}$ are (possibly empty) tuples of \emph{variables}, $p_0,\ldots,p_k$ are \emph{predicate symbols}, $p_0(\tuplevar{x_0})$ is the \emph{head} of the CHC and $p_1(\tuplevar{x_1}) \wedge \ldots \wedge p_k(\tuplevar{x_k}) \wedge \phi$ is the \emph{body}.  Formulas of the form $p(\tuplevar{x})$ are called atomic formulas or simply \emph{atoms}.
A CHC is often written as $p_0(\tuplevar{x_0}) \leftarrow \phi , p_1(\tuplevar{x_1}) , \ldots , p_k(\tuplevar{x_k})$ in the style of constraint logic programs, or \texttt{p0(X0) $\leftarrow$ C, p1(X1),....,pk(Xk)} in text form, where \texttt{C} is a constraint formula.
A \emph{constrained fact} is a CHC $p(\tuplevar{x}) \leftarrow \phi$, where $\phi$ is a constraint over $\tuplevar{x}$.  

\begin{definition}[Ordering on sets of constrained facts]\label{def-prec}

We extend the relation $\preceq$ to sets of constrained facts. Let $S, S'$ be sets of constrained facts. Then $S \preceq S'$ if for each constrained fact 
$p(\tuplevar{x}) \leftarrow \phi$ in $S$ there exists a constrained fact (with variables suitably renamed) $p(\tuplevar{x}) \leftarrow \psi$ in $S'$, such that $\phi \preceq \psi$.  Furthermore, if  $\renames$ is a generalisation operator on constraints then
$S \preceq \{p(\tuplevar{x}) \leftarrow \renames(\phi) \mid p(\tuplevar{x}) \leftarrow \phi \in S\}$.
\end{definition}

We do not go into detail on the translation of imperative programs to CHCs, but note that a distinct predicate symbol is generated for each program point, and the arguments of the predicate for a given program point are the values of the program variables at that point. 
The CHCs defining a predicate capture the transitions in an operational semantics.
Figure \ref{fig-chcs} illustrates the translation of the programs in Figures \ref{fig-pe-exp}(a) and \ref{fig-polyvariance}(a) into CHCs.
Depending on the style of semantic specification (such as small-step semantics or big-step semantics), different CHCs can be obtained for a program.

\begin{figure}[h]
\begin{center}
\begin{tabular}{l|l}

\begin{lstlisting}
start $\leftarrow$ 
   p0(X,Y,Z).
p0(X,Y,Z) $\leftarrow$ 
   Z1=1, 
   while0(X,Y,Z1).
while0(X,Y,Z) $\leftarrow$ 
   Y>0, Z1=X*Z, Y1=Y-1,
   while0(X,Y1,Z1).
while0(X,Y,Z) $\leftarrow$ 
   Y=<0.
\end{lstlisting}
~~~~~&~~~~~
\begin{lstlisting}
start $\leftarrow$ 
   while0(X,Y,M).
while0(X,Y,M) $\leftarrow$
   X>0,
   if0(X,Y,M).
while0(X,Y,M) $\leftarrow$
   X=<0.
if0(X,Y,M) $\leftarrow$
   Y<M, Y1=Y+1, 
   while0(X,Y1,M).
if0(X,Y,M) $\leftarrow$
   Y>=M, X1=X-1, 
   while0(X1,Y,M).
\end{lstlisting}\\
&\\
~~~~~~~~~(a)~~~~~~~&~~~~~~~~~(b)~~~~~~~~\\
\end{tabular}
\end{center}
\caption{CHC representation of (a) Figure \ref{fig-pe-exp}(a) and (b) Figure \ref{fig-polyvariance}(a) }
\label{fig-chcs}
\end{figure}

\subsection{Constrained Horn clause derivations} 
The definitions of CHC derivations and partial evaluation are based on standard definitions (e.g. \cite{Lloyd-Shepherdson-91})
adapted to include constraints, and using the ``resultant" style of derivation.  Instead of an initial query or goal $\leftarrow \phi,p(\tuplevar{x})$,
we start a derivation with a CHC $p(\tuplevar{x}) \leftarrow \phi, p(\tuplevar{x})$ and each step replaces a body literal.
\begin{definition}[Derivation step]
A CHC \emph{derivation step} or \emph{unfolding step} is defined as follows.  Let $c_1,c_2$ be CHCs, where
$c_1 = q_0(\tuplevar{x_0}) \leftarrow \phi , q_1(\tuplevar{x_1}) , \ldots , q_k(\tuplevar{x_k})$ and
$c_2 = q_i(\tuplevar{y_0}) \leftarrow \phi' , r_1(\tuplevar{y_1}) , \ldots , r_m(\tuplevar{y_m})$,
with variables of $c_1$ and $c_2$ renamed apart.
Then the result of \emph{unfolding $c_1$ with $c_2$ on $q_i(\tuplevar{x_i})$} is:

\[
\begin{array}{l}
\begin{cases}
 
\begin{array}{lll}
q_0(\tuplevar{x_0}) &\leftarrow &\phi \wedge \phi' \wedge \tuplevar{x_i}=\tuplevar{y_0},\\
&& q_1(\tuplevar{x_1}) , \ldots , q_{i-1}(\tuplevar{x_{i-1}}), \\
&& r_1(\tuplevar{y_1}) , \ldots , r_m(\tuplevar{y_m}), \\
&& q_{i+1}(\tuplevar{x_{i+1}}), \ldots,q_k(\tuplevar{x_k})
\end{array}
&
\text{if } \SAT(\phi \wedge \phi' \wedge \tuplevar{x_i}=\tuplevar{y_0})\\
q_0(\tuplevar{x_0}) \leftarrow \false & \text{otherwise}
\end{cases}
\end{array}
\]
\end{definition}

\begin{definition}[Derivation tree]\label{derivation-tree}
Let $P$ be a set of CHCs and let $A = p(\tuplevar{x}) \leftarrow \phi$ be a constrained fact where $\phi$ is a constraint on $\tuplevar{x}$.  
Then a \emph{derivation tree} for $A$ in $P$ is a tree where every node is labelled by a CHC, such that:
\begin{itemize}
\item
the root is labelled with $p(\tuplevar{x}) \leftarrow \phi, p(\tuplevar{x})$;
\item
for a non-leaf node labelled with a CHC $c$, its children are labelled with CHCs $\{c_1,\ldots,c_k\}$, 
where $q_i(\tuplevar{x_i})$ is some atom in the body of $c$, $\{d_1,\ldots,d_k\}$ is the set of CHCs
(with variables suitably renamed) in $P$, whose head has predicate $q_i$, and $c_i$ is the result of
unfolding $c$ with $d_i$ on $q_i(\tuplevar{x_i})$;
\item
a leaf node is labelled by a CHC of the form $p(\tuplevar{x}) \leftarrow \varphi$, where $\varphi$ is a constraint (possibly $\false$).
\end{itemize}
The definition is non-deterministic since any atom $q_i(\tuplevar{x_i})$ can be
selected at a non-root node.

\end{definition}

A derivation tree can contain infinite branches. The specialisation algorithm developed in the next section depends on  constructing  a \emph{partial}  derivation tree, which is
a finite tree following Definition \ref{derivation-tree} with the additional case that a leaf node may be labelled by a
CHC of the form $ p(\tuplevar{x_0}) \leftarrow \phi , q_1(\tuplevar{x_1}) , \ldots , q_k(\tuplevar{x_k})$ ($k>0$) representing an
incomplete branch of the derivation tree.
A branch of  a (partial) derivation tree is a \emph{failing} branch if it ends in a leaf labelled by 
$p(\tuplevar{x}) \leftarrow \false$, otherwise it is \emph{non-failing}.

\begin{definition}[Partial evaluation of a constrained fact]\label{def-pe}

Let $A$ be a constrained fact and $P$ be a set of CHCs. Let $T$ be a derivation tree for $A$ in $P$.
Then a \emph{partial evaluation of $A$ in $P$} is a finite set of CHCs $\{c_1,\ldots,c_m\}$ labelling nodes chosen from
the non-root nodes of $T$ such that there is exactly one node for each non-failing branch of $T$.

\end{definition}

Clearly, the whole derivation tree does not have to be constructed in order to get a partial evaluation, but only an initial portion;
then one CHC from each
branch is collected.

The non-determinism in the definition of a derivation tree is resolved by an \emph{unfolding rule}, which both selects which body atom
to unfold at each step, and decides when to stop extending a branch in a (partial) derivation tree in order to return a partial evaluation.

\begin{definition}[Unfolding rule]
An unfolding rule $U$ is a function which given a set of CHCs $P$ and a constrained fact $A$, returns exactly one finite set
of CHCs that is a partial evaluation of $A$ in $P$. For a set $S$ of constrained facts, the set of CHCs obtained by applying $U$ to each element of $S$ is called a partial
evaluation of $S$ in $P$ using $U$.
\end{definition}

\begin{example}
Let $P$ be the set of clauses in Figure \ref{fig-chcs}(b).  Let $A$ be the constrained atom $\mathtt{while0(X,Y,M)} \leftarrow \true$.
Figure \ref{fig-pe-ex} shows three sets of CHCs that are examples of partial evaluations of $A$ in $P$, with different unfolding rules.  Note that for CHCs with at most one atom in the body, such as in this example, the choice of atom for each derivation step is determinate, so the unfolding rule only determines how far to unfold each branch.

\begin{figure}
\begin{tabular}{l|l|l}
\begin{lstlisting}
while0(X,Y,M) $\leftarrow$
   X>0,
   if0(X,Y,M).
while0(X,Y,M) $\leftarrow$
   X=<0.
\end{lstlisting}
~~&~~
\begin{lstlisting}
while0(X,Y,M) $\leftarrow$
   X>0,
   Y<M, Y1=Y+1, 
   while0(X,Y1,M).
while0(X,Y,M) $\leftarrow$
   X>0,
   Y>=M, X1=X-1, 
   while0(X1,Y,M).
while0(X,Y,M) $\leftarrow$
   X=<0.
\end{lstlisting}
~~&~~
\begin{lstlisting}
while0(X,Y,M) $\leftarrow$
   X>0,
   Y<M, Y1=Y+1, 
   X>0, 
   if0(X,Y1,M).
while0(X,Y,M) $\leftarrow$
   X>0,
   Y<M, Y1=Y+1, 
   X=<0.
while0(X,Y,M) $\leftarrow$
   X>0,
   Y>=M, X1=X-1, 
   while0(X1,Y,M).
while0(X,Y,M) $\leftarrow$
   X=<0.
\end{lstlisting}
\end{tabular}
\caption{Three possible partial evaluations of constrained fact $\mathtt{while0(X,Y,M)} \leftarrow \true$ in Figure \ref{fig-chcs}(b).
The leftmost column is the trivial unfolding, consisting of the original clauses for $\mathtt{while0}$.}\label{fig-pe-ex}
\end{figure}
\end{example}

\section{Specialisation algorithm}
Algorithm \ref{alg:pe} shows an outline specialisation algorithm SP for CHCs based on the
``basic algorithm" in \cite{gallagher:pepm93}.
This is a so-called \emph{online} specialisation algorithm, which makes control decisions on evaluation and polyvariance
on the fly, 
as opposed to
\emph{offline} specialisation, in which control decisions are determined by the results of a prior analysis such 
as a binding time analysis.

The algorithm takes as input a set of CHCs $P$ and a set of entry points $S_0$, where each entry point is a constrained fact.
It is parameterised by two operations, namely 
$\unfold$ and $\abst_\renames$.  
\begin{itemize}
\item
$\unfold(S)$ returns a partial evaluation (Definition \ref{def-pe}) of the set of constrained facts $S$.
\item 
$\abst_\renames(S)$ is a set of constrained facts such that 
$\abst_\renames(S) = \{p(\tuplevar{x}) \leftarrow \renames(\phi) \mid p(\tuplevar{x}) \leftarrow \phi \in S\}$, where
$\renames$ is some generalisation operator (Definition \ref{def-gen}).
\end{itemize}
Two other functions are called in the algorithm, $\bodyfacts$ and $\rename$.
\begin{itemize}
\item
$\bodyfacts(Q)$ returns the set of constrained atoms collected from the bodies of clauses in a set of CHCs $Q$.
It is defined as $\bodyfacts(Q) = \{ p_i(\tuplevar{x_i}) \leftarrow \phi\vert_{\tuplevar{x_i}} \mid 
p_0(\tuplevar{x_0}) \leftarrow \phi , p_1(\tuplevar{x_1}) , \ldots , p_k(\tuplevar{x_k}) \in Q\}$.

\end{itemize}
The $\rename$ function is an unfold-fold transformation, which will be discussed in Section \ref{correctness} and in Example \ref{pe-alg-ex}.

\begin{algorithm}[h!]
\begin{algorithmic}[1]
     \State \textbf{Input:} Finite set of CHCs $P$,  finite set of constrained facts $S_0$, generalisation operator $\renames$.
     \State \textbf{Output:} Finite set of CHCs
      \State $S \gets S_0$
      \Repeat 
      	   \State $S'=S$
           \State $S \gets S \cup \abst_\renames(\bodyfacts(\unfold(S)))$
       \Until $S'=S$
       \State \Return $\rename(S)$
\end{algorithmic}
\caption{SP($P$,$S_0$)\label{alg:pe}}
\end{algorithm}

The successive values of $S'$ in the \textbf{repeat} loop of the algorithm (line 7) form an increasing sequence of sets of
constrained facts with respect to $\preceq$ starting from the input
set $S_0$, say $S_0,S_1,S_2,\ldots$; the loop terminates
if for some $j > 0$, $S_{j-1} = S_j$.

\subsection{Correctness of the specialisation algorithm}\label{correctness}
The loop termination condition on line 7 of Algorithm \ref{alg:pe} establishes a \emph{closedness} condition. 
\begin{lemma}\label{closedness}
Let $S$ be the final set of constrained facts when the loop terminates.  Then 
$\bodyfacts(\unfold(S))  \preceq S$.
\end{lemma}
The proof of the lemma relies directly on the fact that $\abst_\renames$ in line 6 uses a generalisation operator (Definition \ref{def-gen}).
Lemma \ref{closedness} is the basis for a proof of correctness using unfold-fold proofs \cite{Pettorossi-Proietti,EtG96,DeAngelisFPP18}.  
Line 8 of the algorithm is defined as an \emph{unfold-fold proof}, where folding is performed 
using a set of new definitions constructed from the set $S$ obtained from the loop. Specifically,
$\Def_S = \{p'(\tuplevar{x}) \leftarrow \phi, p(\tuplevar{x}) \mid p(\tuplevar{x}) \leftarrow \phi \in S\}$, where $p'$ is a fresh predicate symbol unique for each element of $S$. These definitions are unfolded, and then folded using $\Def_S$.
Lemma \ref{closedness} guarantees that all the predicates from the input clauses can be folded to their renamed versions 
in $\Def_S$.
Thus in the final transformed program, every predicate in the head of a clause in 
$\Def_S$ calls only head predicates in $\Def_S$. The clauses returned by $\rename(S)$are just those defining head predicates in $\Def_S$, since the original predicates are unreachable from initial set of queries $S_0$.  
(In practice, we can then rename versions from $S_0$ back to their original predicate names.)

In short, the main loop of Algorithm \ref{alg:pe} constructs a set of new definitions $\Def_S$, 
while correctness of the clauses returned by the algorithm
follows from the general results on unfold-fold transformations using $\Def_S$, along with the closeness property of Lemma \ref{closedness}.

\section{Property-based abstraction}

We now turn to the consideration of the abstraction function $\abst_{\renames}$, using property-based abstraction.
We first define a generalisation operator $\renames_\Psi$.
Let $\Psi \subseteq \C^T$ be a finite set of constraints. Given a formula $\phi \in \C^T$, then 
\[
\renames_\Psi(\phi) = \bigwedge\{ \psi \mid \psi \in \Psi ,\phi \preceq \psi\} 
\wedge \bigwedge\{ \neg\psi \mid \psi \in \Psi, \phi \preceq \neg\psi\}.
\]

\begin{lemma}\label{lemma-rep}
$\renames_\Psi$ is a generalisation operator, that is, for all $\phi \in \C^T$, $\phi \preceq_T \renames_\Psi(\phi)$.
\end{lemma}
Note that $\renames_\Psi(\phi)$ is a conjunction of elements of $\Psi$ and negations of elements of $\Psi$. Since we assume that
$\Psi$ is finite, then the set of possible values of $\renames_\Psi(\phi)$ is finite.
If none of the elements of $\Psi$ or their negations is entailed by $\phi$ then $\renames_\Psi(\phi) = \true$.

\begin{example}\label{ex2}
Consider a set $\Psi = \{\psi_1,\psi_2,\psi_3\}$, where $\psi_1 \wedge \psi_3 = \false$, 
 $\psi_2 \wedge \psi_3 = \false$ and $\psi_1 \wedge \psi_2 \neq \false$. Figure \ref{fig-ex1} shows the generalisation for various choices of a property $\varphi$, that is, the values of $\renames_\Psi(\varphi)$.
\begin{figure}
\begin{center}
\includegraphics[width=0.50\textwidth]{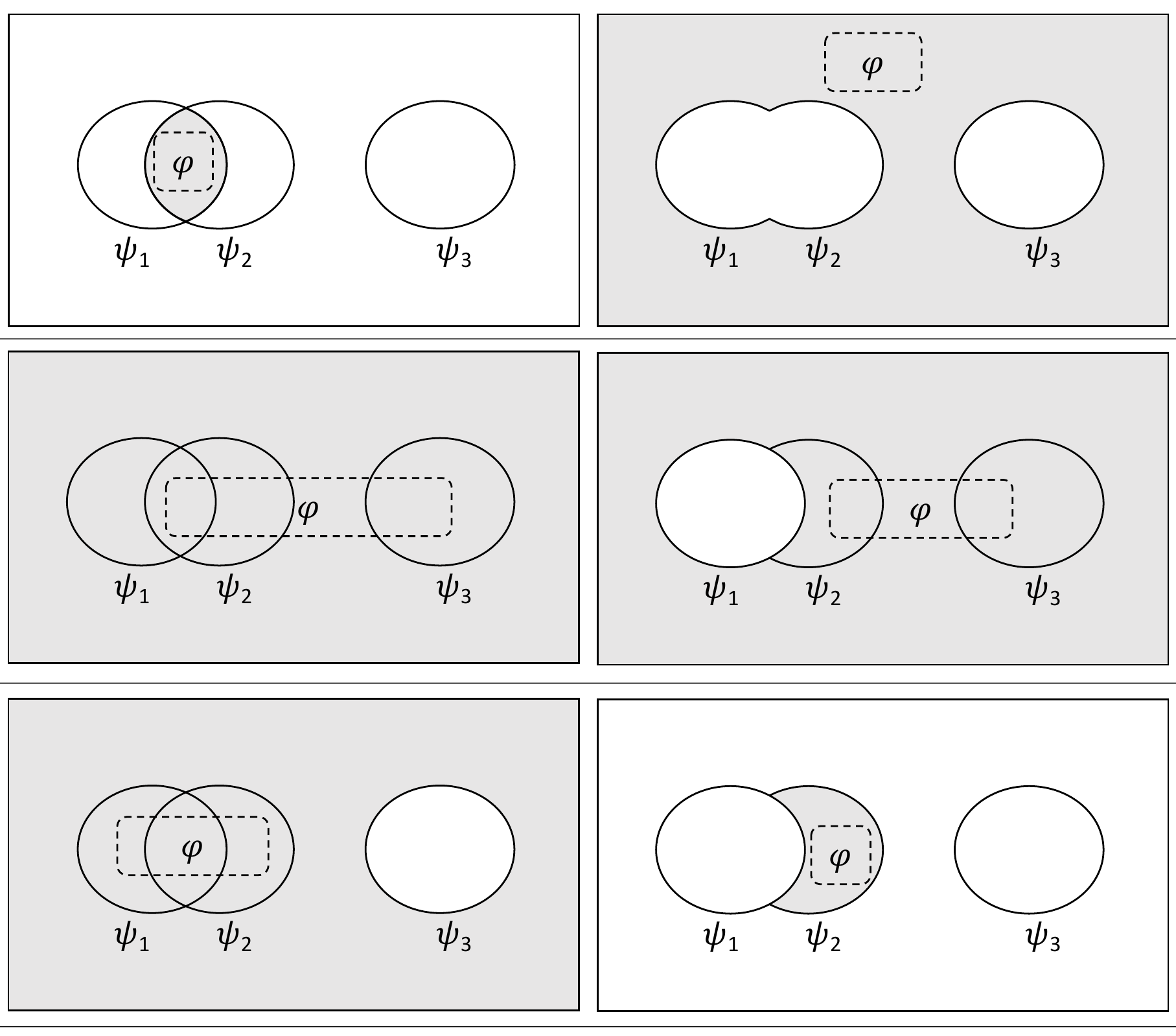}
\end{center}
\caption{The property to be generalised ($\varphi$) is shown as an area with dotted outline, and the shaded areas show the generalisation of $\varphi$ using operation $\renames_\Psi(\varphi)$, where $\Psi = \{\psi_1,\psi_2,\psi_3\}$.}
\protect\label{fig-ex1}
\end{figure}

\end{example}
$\renames_\Psi$ is extended to apply to constrained facts, and for convenience we take the set $\Psi$ also to consist of constrained facts.
\[
\renames_\Psi(p(\tuplevar{x}) \leftarrow \phi) = p(\tuplevar{x}) \leftarrow \bigwedge\{ \psi \mid p(\tuplevar{x}) \leftarrow\psi \in \Psi ,\phi \preceq \psi\} 
\wedge \bigwedge\{ \neg\psi \mid p(\tuplevar{x}) \leftarrow\psi \in \Psi, \phi \preceq \neg\psi\}.
\]

 Let $S$ and $\Psi$ be 
sets of constrained facts. The operation $\abst_{\renames}$ from Algorithm \ref{alg:pe} is defined where $\renames$ is the generalisation operator $\renames_\Psi$.
\[
\abst_{\renames_\Psi}(S) = \{p(\tuplevar{x}) \leftarrow \renames_\Psi(\varphi) \mid p(\tuplevar{x}) \leftarrow \varphi \in S\}
\]
We now have all the components of the algorithm, and we show an example of specialisation with property-based abstraction. 

\begin{example}\label{pe-alg-ex}
Let $P$ be the set of CHCs representing the code
in Figure \ref{fig-polyvariance}(a), that is, the clauses from Figure \ref{fig-chcs}(b).  Let $S_0 = \{\mathtt{start} \leftarrow \true\}$ and 
let the set $\Psi$ contain the following constrained facts.

\begin{tabular}{lll}
\begin{lstlisting}
  while0(A,B,C) $\leftarrow$  A>0
  while0(A,B,C) $\leftarrow$ A$\le$0
\end{lstlisting}&~~~
\begin{lstlisting}
  while0(A,B,C) $\leftarrow$  B<C
  while0(A,B,C) $\leftarrow$ B$\ge$C
\end{lstlisting}&~~~
\begin{lstlisting}
if0(A,B,C) $\leftarrow$ B<C
if0(A,B,C) $\leftarrow$ B$\ge$C
\end{lstlisting}\\
\end{tabular}

The following unfolding rule is used: a clause body is unfolded until either a branch point is reached (i.e. a call to a predicate 
that appears in the head of more than one clause) or a recursive predicate is reached (i.e. a predicate that is the target of a back edge in the predicate dependency graph of the program traversed from the initial predicate \texttt{start}).  In the given clauses, this implies that each partial evaluation consists of only one unfolding step since every predicate is either recursive or a branch.
Algorithm \ref{alg:pe} proceeds as follows, with $S_0$ initialised to $\{\mathtt{start} \leftarrow \true\}$.

\begin{itemize}
\item
Iteration 1: $S_1 = S_0 \cup \{\mathtt{while0(A,B,C)} \leftarrow \true\}$.
\item
Iteration 2: $S_2 = S_1 \cup \{\mathtt{if0(A,B,C)} \leftarrow \mathtt{A>0}\}$.
\item
Iteration 3: $S_3 = S_2 \cup \{\mathtt{while0(A,B,C)} \leftarrow \mathtt{A>0},\ \ \  \mathtt{while0(A,B,C)} \leftarrow \mathtt{B \ge C}\}$.
\item
Iteration 4: $S_4 = S_3 \cup \{\mathtt{if0(A,B,C)} \leftarrow \mathtt{A>0,B\ge C}\}$.
\item
Iteration 5: $S_5 = S_4$.
\end{itemize}
In this sequence we applied the operator $\abst_{\renames_\Psi}$ at each stage to reach the sets shown in the sequence.
(To be precise, we apply the generalisation operator only to constrained facts for the recursive predicate $\mathtt{while0}$, which is sufficient to ensure termination).
For example, in iteration 3, partial evaluation of the constrained fact $\mathtt{if0(A,B,C)} \leftarrow \mathtt{A>0}$
results in the following CHCs.
\begin{lstlisting}
if0(A,B,C) $\leftarrow$ A>0,C>B,D=B-1,while0(A,D,C).
if0(A,B,C) $\leftarrow$ A>0,B$\ge$C,D=A-1,while0(D,B,C).
\end{lstlisting}
The constraints projected onto the body atoms $\mathtt{while0(A,D,C)}$ and $\mathtt{while0(D,B,C)}$ are respectively
$\mathtt{A>0}$ and 
$\mathtt{D> -1}, \mathtt{B\ge C}$. The result of applying abstraction, that is, $$\abst_{\renames_\Psi}(\{\mathtt{while0(A,D,C) \leftarrow A>0}, \ \ \ \ 
\mathtt{while0(D,B,C) \leftarrow D> -1,B\ge C}\})$$
is $\{\mathtt{while0(A,B,C) \leftarrow A>0, 
while0(A,B,C) \leftarrow B\ge C}\}$.
Notice that the second constrained fact has been generalised; evaluating $\renames_\Psi(\mathtt{while0(D,B,C) \leftarrow D> -1,B\ge C}\})$ yields $\mathtt{while0(A,B,C) \leftarrow B\ge C}$, since $\mathtt{B\ge C}$ is the only element of $\Psi$ that is entailed.
Similarly, in iteration 5, partial evaluation of $\mathtt{if0(A,B,C) \leftarrow A>0,B\ge C}$ 
results in the CHC:
\begin{lstlisting}
if0(A,B,C) $\leftarrow$ A>0,B$\ge$C,D=A-1,while0(D,B,C).
\end{lstlisting}
The constraint on the body atom $\mathtt{while0(D,B,C)}$ is $\mathtt{D> -1,B\ge C}$;  as before this is generalised to
the constrained fact
$\mathtt{while0(A,B,C) \leftarrow B\ge C}$.

The function $\rename_\renames$ of Algorithm \ref{alg:pe} constructs a set of renaming definitions, as follows.
\[
\begin{array}{l|l}
\mathtt{start} \leftarrow \mathtt{start},\true &
~~~\mathtt{if0_1(A,B,C)} \leftarrow \mathtt{if0(A,B,C)},  \mathtt{A>0,B\ge C} \\
\mathtt{while0_2(A,B,C)} \leftarrow \mathtt{while0(A,B,C)},  \mathtt{B \ge C}~~~ & 
~~~\mathtt{while0_3(A,B,C)} \leftarrow \mathtt{while0(A,B,C)}, \mathtt{A>0}  \\
\mathtt{if0_4(A,B,C)} \leftarrow \mathtt{if0(A,B,C)} , \mathtt{A>0}  &
~~~\mathtt{while0_5(A,B,C)} \leftarrow \mathtt{while0(A,B,C)},  \true \\
\end{array}
\]
These definitions are unfolded (using the same strategy as for the $\unfold$ operation) and the atoms in the bodies of the 
unfolded clauses are folded using the above clauses.
This gives the following specialised CHC clauses.
\begin{lstlisting}
start $\leftarrow$ while0$_5$(A,B,C).
while0$_5$(A,B,C) $\leftarrow$ A>0,if0$_4$(A,B,C).
while0$_5$(A,B,C) $\leftarrow$ -A>=0.
if0$_4$(A,B,C) $\leftarrow$ A>0,-B+C>0,B-D= -1,while0$_3$(A,D,C).
if0$_4$(A,B,C) $\leftarrow$ A>0,B-C>=0,A-D=1,while0$_2$(D,B,C).
while0$_3$(A,B,C) $\leftarrow$ A>0,if0$_4$(A,B,C).
while0$_2$(A,B,C) $\leftarrow$ B-C>=0,A>0,if0$_1$(A,B,C).
while0$_2$(A,B,C) $\leftarrow$ B-C>=0,-A>=0.
if0$_1$(A,B,C) $\leftarrow$ A>0,B-C>=0,A-D=1,while0$_2$(D,B,C).
\end{lstlisting}
The predicate dependency graph for these clauses is shown in Figure \ref{fig-cfg-ex3} and it can be seen that this has the same structure as the code in Figure \ref{fig-polyvariance}(b) where there are two distinct loops.  Note that the predicate $\mathtt{while0_5}$ is in fact not the head of a loop but rather the initial if-statement of Figure \ref{fig-polyvariance}(b), while the predicate $\mathtt{if0_4}$
is the head of a loop.
\begin{figure}
\begin{center}
\includegraphics[width=0.3\textwidth]{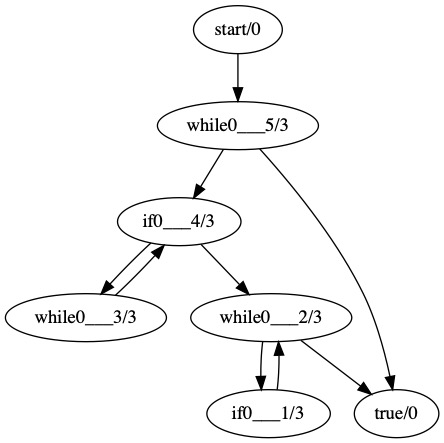}
\end{center}
\caption{The predicate dependency graph for Example \ref{pe-alg-ex}}
\protect\label{fig-cfg-ex3}
\end{figure}
\end{example}

\section{Choice of properties and granularity of abstraction}

The set of properties $\Psi$ in Example \ref{pe-alg-ex} was chosen so that the properties were relevant to the tests determining the control flow. However, the choice of properties can be critical to achieving good specialisations.
In this section we discuss
the effect of the choice of properties. 

In general, it is clear that the larger the set of properties, the more versions of predicates can be produced, and thus more specialised
clauses can be generated. Fewer properties, on the other hand, cause information needed for specialisation to be lost.
 For example, the following properties could also be chosen in Example \ref{pe-alg-ex}. It is a subset of the set previously chosen, incorporating only the constraints directly appearing in the clauses for the respective 
predicates.
\begin{lstlisting}
{ while0(A,B,C) $\leftarrow$  A>0,   while0(A,B,C) $\leftarrow$ A$\le$0, 
  if0(A,B,C) $\leftarrow$ B<C,  if0(A,B,C) $\leftarrow$ B$\ge$C }
\end{lstlisting}
Using this choice of $\Psi$, no specialisation at all is achieved; the original clauses are returned. The 
problem is that the constraints on $\mathtt{if0}$ are lost when abstracting the calls to $\mathtt{while0}$, since the properties applying 
to $\mathtt{while0(A,B,C)}$ say nothing about the values of $\mathtt{B}$ or $\mathtt{C}$.

However, for a given unfolding rule, there is a limit to how much specialisation can be achieved, no matter how many properties
$\Psi$ contains.  Consider the following set $\Psi$ for Example \ref{pe-alg-ex}, which results from collecting all constraints from the given clauses, projected onto head and body atoms.

\begin{tabular}{lll}
\newline
\begin{lstlisting}
while0(A,B,C) $\leftarrow$ A>0
while0(A,B,C) $\leftarrow$ A$\le$0
\end{lstlisting}
&~~~~
\begin{lstlisting}
while0(A,B,C) $\leftarrow$ C>B-1
while0(A,B,C) $\leftarrow$ B$\ge$C
\end{lstlisting}
&~~~~
\begin{lstlisting}
if0(A,B,C) $\leftarrow$ A>0
if0(A,B,C) $\leftarrow$ B<C
if0(A,B,C) $\leftarrow$ B$\ge$C
\end{lstlisting}\\
\newline
\end{tabular}

\begin{figure}
\begin{center}
\includegraphics[width=0.25\textwidth]{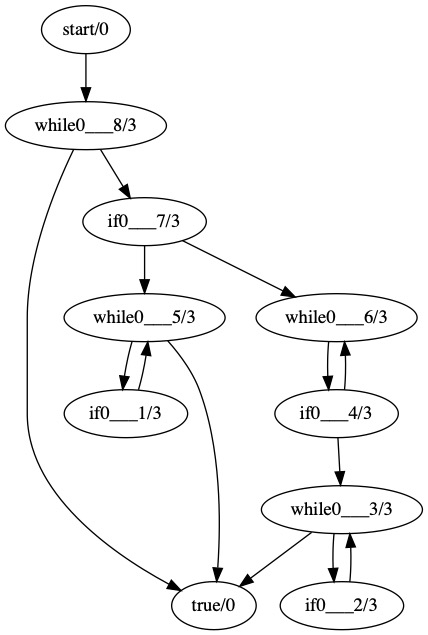}
\end{center}
\caption{The predicate dependency graph for Example \ref{pe-alg-ex}, with enlarged set of properties.}
\protect\label{fig-cfg3-ex3}
\end{figure}

Figure \ref{fig-cfg3-ex3} shows the predicate dependency graph for the clauses resulting from this set.
While there are more versions of predicates than in Figure \ref{fig-cfg-ex3}, the corresponding clauses are no more
specialised.  Viewing the graph as a finite automaton, it can be verified that the states $\mathtt{if0\_4}$ and $\mathtt{if0\_7}$ 
are equivalent, as are $\mathtt{while0\_5}$ and $\mathtt{while0\_3}$, and $\mathtt{if0\_1}$ and $\mathtt{if0\_2}$.  In short, the automaton can be minimised to give the same automaton as in Figure \ref{fig-cfg-ex3}.  
For this example, there can be no
better specialisation for the input clauses in Example \ref{pe-alg-ex}, with any set of properties $\Psi$ or indeed any
other unfolding rule, than the one achieved in Example \ref{pe-alg-ex}.
Space does not permit a detailed account of the
automata-theoretic argument, but we state the following conjecture.

\begin{proposition}\label{prop1}
Let $P$ be a set of CHCs and $S_0$ a set of constrained facts, $\Psi$ a set of properties and $U$ an unfolding rule. Let $P'$ be the output of  Algorithm \ref{alg:pe} with $P$ and $S_0$ as inputs, using $\Psi$ and $U$. Let  $\{p_1,\ldots,p_k\}$ be the set of predicates from $P$ of which the predicates of $P'$ are variants. Then there exists a set of clauses $P''$ having a minimal number of variants of the same set $\{p_1,\ldots,p_k\}$, that is equivalent to $P'$ wrt to derivations starting with $S_0$. 
\end{proposition}
Proof sketch: $P''$ can be constructed by minimisation of a tree automaton derived from $P'$.  
Note that $\{p_1,\ldots,p_k\}$ could be a strict subset of the predicates of $P$ and so $P$ itself does not in general 
satisfy the condition for $P''$.   Although $P''$ is minimal 
in the sense given above, 
further specialisation might be achievable using a different unfolding rule than $U$. We also conjecture that there exists a set of properties $\Psi'$ such that
executing  Algorithm \ref{alg:pe} with $\Psi'$, $U$, $P$ and $S_0$ would yield $P''$.

\section{Polyvariant specialisation in verification}\label{applications}

In this section we discuss
the role of polyvariant specialisation of CHCs in program verification tasks. 

\paragraph{Pre-condition inference.}
In
\cite{DBLP:journals/tplp/KafleGGS18}, Algorithm \ref{alg:pe} was used as a component in an algorithm for computing 
sufficient conditions for safety of imperative programs encoded as CHCs.  In many cases, the    safety condition is 
a disjunction.  Polyvariant specialisation enabled the relevant disjuncts to be found by a convex polyhedral analysis.  

\begin{example}\label{ex-tplp}
Consider the example in Figure \ref{fig-ex-tplp} taken from \cite{DBLP:journals/tplp/KafleGGS18}.  Note that the translation to 
CHCs corresponds to a backwards flow of control from the error predicate $\mathtt{false}$ to the program start predicate $\mathtt{init}$.
The goal is to infer conditions on the start predicate that ensure that the error predicate is not reached.
\begin{figure}[t]
  \begin{center}
  \begin{tabular}{l|l}
    \begin{lstlisting}
    int a, b; 
    if (a $\leq$ 100) 
        a = 100-a;    
    else
        a=a-100;
    while (a $\geq$ 1)     
      	a=a-1; 
      	b=b-2;
    $\pmb{\mathtt{assert}}$(b != 0);
    \end{lstlisting}
    &
    \begin{lstlisting}
   init(A,B) $\leftarrow$ true.
   if(A,B) $\leftarrow$ A0 $\leq$ 100, A=100-A0, init(A0,B).
   if(A,B) $\leftarrow$ A0 $\geq$ 101, A=A0-100, init(A0,B).
   while(A,B) $\leftarrow$ if(A,B).
   while(A,B) $\leftarrow$ A0$\geq$1, A=A0-1, B=B0-2, 
                  while(A0,B0).
   false $\leftarrow$ A$\leq$0, B=0, 
                  while(A,B)
   \end{lstlisting}
  \end{tabular}
  \end{center}
  \caption{Example from \cite{DBLP:journals/tplp/KafleGGS18}: (left) original program, (right) translation to
   CHCs}
 \label{fig-ex-tplp}
\end{figure}
Specialisation of the set of CHCs was carried out using the following set of properties, and the same unfolding rule as in
Example \ref{pe-alg-ex}.
\begin{lstlisting}
{if(A,B) $\leftarrow$ A$\ge$0,   if(A,B) $\leftarrow$ A$\ge$1, 
init(A,B) $\leftarrow$ A$\ge$101, init(A,B) $\leftarrow$ -A$\ge$ -100,
while(A,B) $\leftarrow$ A$\ge$0, while(A,B) $\leftarrow$ A$\ge$1, while(A,B) $\leftarrow$ -A$\ge$0,B=0,
while(A,B) $\leftarrow$ -A$\ge$0, while(A,B) $\leftarrow$ B=0}
\end{lstlisting}
\begin{figure}
\begin{center}
\begin{tabular}{ll}
\includegraphics[width=0.09\textwidth]{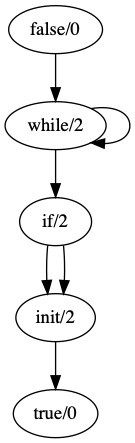}
&~~~~
\includegraphics[width=0.25\textwidth]{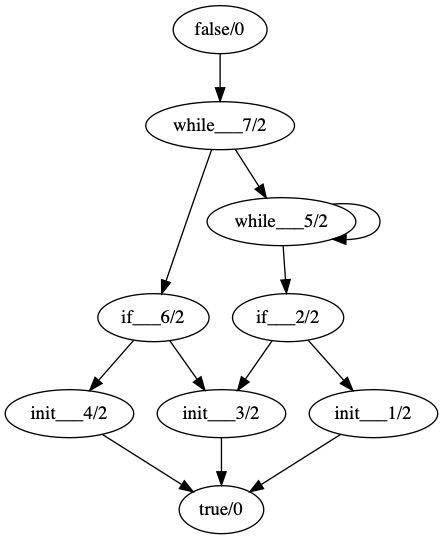}
\end{tabular}
\end{center}
\caption{The predicate dependency graph for Example \ref{ex-tplp}, before and after polyvariant specialisation}
\protect\label{fig-tplp-ex}
\end{figure}
It can be seen that three versions of the predicate $\mathtt{init}$ have been generated, arising from different paths through
the program.  Analysis of these specialised CHCs allowed the disjunctive precondition on $\mathtt{init(A,B)}$, namely $\mathtt{B \neq |2A-100|}$ to be derived. This condition could not be derived from the original code without an analysis domain of disjunctive properties, which is more difficult to implement and control.  Polyvariant specialisation, in effect, provides a heuristic
for introducing disjunctions selectively, where they can affect the control flow.

\end{example}

\paragraph{Termination analysis.}
In \cite{DomenechGG2018}, polyvariant specialisation was used to transform a control-flow graph obtained from a program 
into another equivalent control-flow graph in which the loops were in a form more suitable for automatic proof of termination.
The control-flow graphs are straightforwardly 
represented as CHCs and examples of their polyvariant specialisation are displayed in \cite{DomenechGG2018}.
The example from Figure \ref{fig-polyvariance} provides a case in point.  The structure of the single loop makes it rather hard 
to find a suitable ranking function that establishes termination; whereas the restructured code based on polyvariant specialisation,
 with two separate loops, is easy to
prove terminating, since each loop has a simple ranking function.  A large number of experiments was reported in
\cite{DomenechGG2018}, showing that polyvariant specialisation of the control-flow graph very often improves the 
effectiveness of both automatic termination analysis and complexity bound analysis. In short, the work demonstrates that control-flow refinement \cite{DBLP:conf/pldi/GulwaniJK09} can be achieved by polyvariant specialisation. Another relevant approach 
for finding better loop invariants
 is the ``splitter predicates" method \cite{DBLP:conf/cav/SharmaDDA11}; this can also be reproduced using property-based specialisation.

\begin{figure}
\begin{center}
\includegraphics[width=0.5\textwidth]{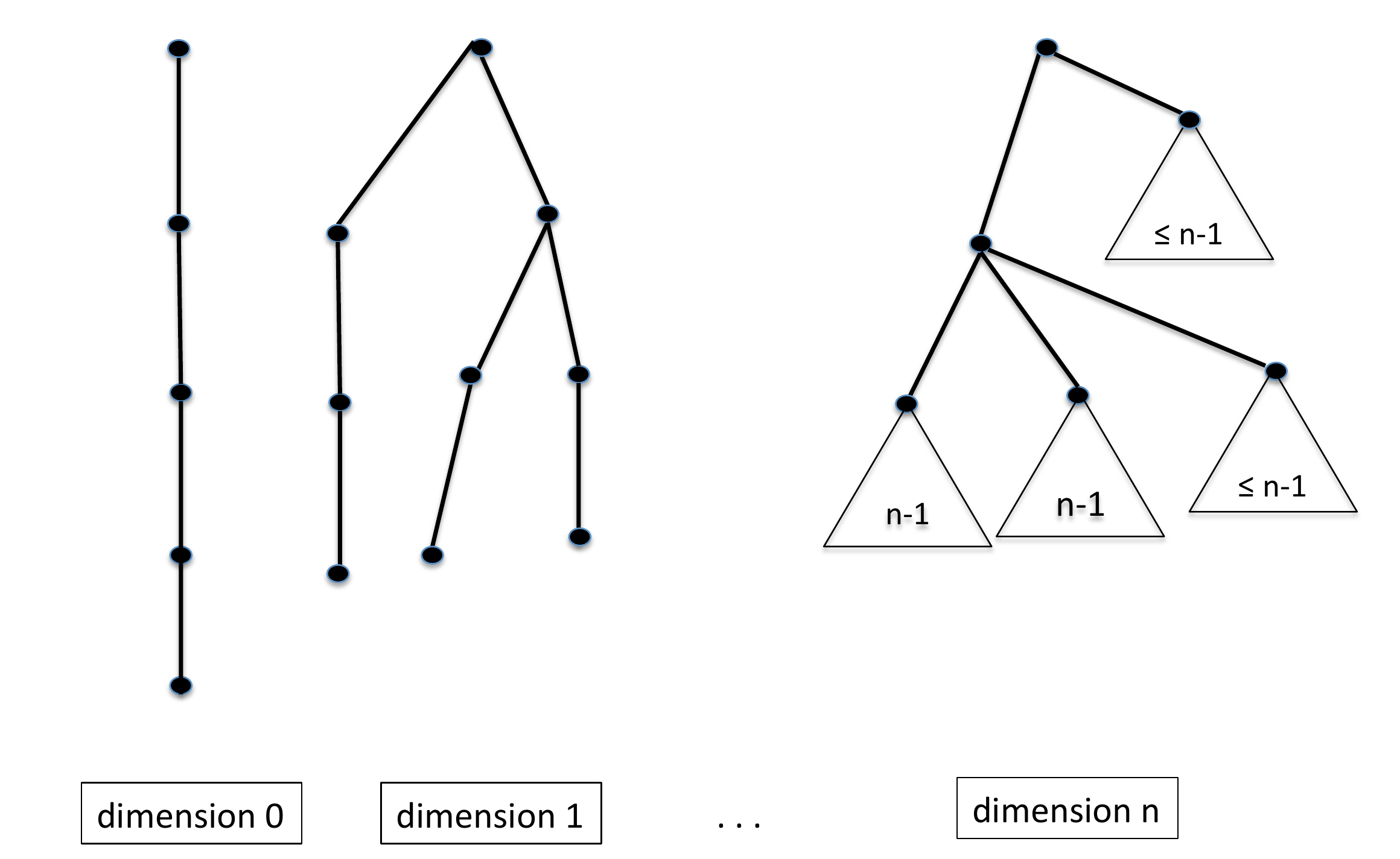}
\end{center}
\caption{The dimension of a node in a tree: leaf nodes have dimension 0; a node has dimension $n+1$ if at least two subtrees have dimension $n$; it has dimension $n$ if exactly one subtree has dimension $n$ and any other subtrees have lower dimension. }
\protect\label{fig-dim}
\end{figure}

\paragraph{Dimension-based decomposition.}
The concept of \emph{tree dimension} has been applied in verification to decompose a 
proof.  The dimension of a set of CHCs is a measure of their non-linearity. A set of CHCs of dimension zero contains only linear clauses (that is, clauses having at most one atom in the body).  Proof trees in such sets of CHCs have no branching.
Sets of clauses of higher dimension give rise to branching proof trees, and the dimension of a tree is determined by the dimensions of the subtrees of the root, as illustrated in Figure \ref{fig-dim}.  A more detailed definition can be found in \cite{DBLP:journals/tplp/KafleGG18}. 
Given a set of CHCs $P$ for which some property is to be verified, we may decompose the problem by dimensions.
For a given dimension $d$, we can define a set of CHCs, say $P^d$, such that an atom $A$ has a proof in $P$ of dimension $d$ if and only if it has a proof in $P^d$.  Since every proof has some finite dimension,  $P \vdash A$ if and only if $P^0 \vdash A \vee P^1 \vdash A \vee P^2 \vdash A \ldots$.    In \cite{DBLP:journals/tplp/KafleGG18}, we showed a technique using 
polyvariant specialisation for generating various dimension-bounded sets of clauses. We could generate $P^d$, the set of clauses yielding exactly the proof trees of dimension $d$; we could also generate $P^{\le d}$, the set of clauses yielding proof trees of dimension at most $d$; and we could generate  $P^{> d}$, the set of clauses yielding proof trees of dimension at least $d+1$.
This enabled a variety of strategies for proof decomposition,  described in detail in \cite{DBLP:journals/tplp/KafleGG18}.

In order to derive such dimension-constrained sets of clauses, we first instrumented the clauses with an extra argument for each
predicate, representing the dimension, together with constraints capturing the rule for computing dimension. That is, the clause
$p_0(\tuplevar{x_0}) \leftarrow \phi , p_1(\tuplevar{x_1}) , \ldots , p_n(\tuplevar{x_n})$ is replaced by
\[
p_0(\tuplevar{x_0},k) \leftarrow \phi , p_1(\tuplevar{x_1},k_1) , \ldots , p_n(\tuplevar{x_n},k_n),dim(k_1,\ldots,k_n,k)
\]
\noindent
where $dim(k_1,\ldots,k_n,k)$ represents the computation of the head dimension $k$ from the (subtree) dimensions $k_1,\ldots,k_n$.

Specialisation was then performed with respect to constraints on $k$. The set of properties $\Psi$ input to
Algorithm \ref{alg:pe} consisted a constrained fact for each dimension up to the required bound. 
\begin{example}\label{ex-fib}
Let $P$ be the set of clauses for the Fibonacci function, instrumented with the dimension as described above, together with a constraint representing a property to be proved and a dimension bound on $\mathtt{false}$ of 2.
\begin{lstlisting}
fib(A,B,0) :- A>=0, A=<1, A=B.
fib(A,B,K) :- A>1, D=A-2, E=A-1, B=F+G, fib(D,G,K2), fib(E,F,K1), 
        K1+1=<K, K2=K.
fib(A,B,K) :- A>1, D=A-2, E=A-1, B=F+G, fib(D,G,K1), fib(E,F,K2), 
        K1+1=<K, K=K2.
fib(A,B,K) :- A>1, D=A-2, E=A-1, B=F+G, fib(D,G,K1), fib(E,F,K2),  
        K1=K-1, K2=K1.
false(A) $\leftarrow$ X>5, fib(X,Y,K), Y<X, K$\le$2.
\end{lstlisting}
Let $\Psi$ be the following set of constrained facts.
\begin{lstlisting}
fib(A,B,C)$\leftarrow$ C$\le$2, fib(A,B,C)$\leftarrow$ C$\le$1, fib(A,B,C)$\leftarrow$ C$\le$0, fib(A,B,C)$\leftarrow$ C$\ge$0,
false(A)$\leftarrow$ A$\le$2, false(A)$\leftarrow$ A$\le$1, false(A)$\leftarrow$ A$\le$0, false(A)$\leftarrow$ A$\ge$0,
\end{lstlisting}
Figure \ref{fig-fib} shows the predicate dependency graphs before and after polyvariant specialisation using initial call $\mathtt{false(A) \leftarrow A \le 2}$, $\Psi$ shown above and an
unfolding rule that just unfolds one step.  Here, $\mathtt{fib\_1}$,  $\mathtt{fib\_2}$ and $\mathtt{fib\_3}$ yield proof trees of dimension 
$\le 0$, $\le 1$ and $\le 2$ respectively.
\begin{figure}
\begin{center}
\begin{tabular}{ll}
\includegraphics[width=0.15\textwidth]{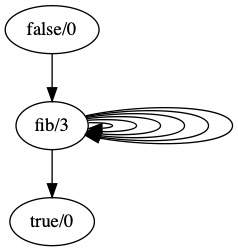}
&
\includegraphics[width=0.12\textwidth]{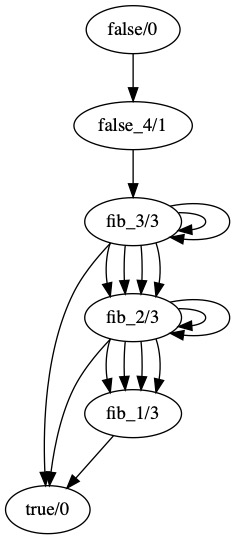}
\end{tabular}
\end{center}
\caption{The predicate dependency graph for Example \ref{ex-fib} producing clauses yielding proof trees of dimension at most 2, before and after polyvariant specialisation}
\protect\label{fig-fib}
\end{figure}
\end{example}
 
\section{Discussion and related work}

The control of partial evaluation, and more broadly of program specialisation, has been much studied \cite{Jones-Gomard-Sestoft}.  The problem arises due to the two termination problems in specialisation: local termination, ensuring that loops in the program are not unfolded indefinitely, and
global termination, which can be seen as the problem of generating only a finite number of versions of each program point.
These two problems are mutually dependent: the more conservative the local unfolding strategy is (in order to ensure local termination), the more important it is to
allow multiple versions of program points in order to preserve information, thus increasing the risk of global non-termination.
In some approaches, the two problems are merged in order to handle these interactions \cite{Turchin-88,DBLP:journals/ngc/Sahlin93,Martens-Gallagher-ICLP95,Leuschel-Martens-96}.  However, from the point of view of conceptual clarity and 
implementability, we argue that it is desirable to separate the two problems.  Thus it becomes important to 
have a flexible way of controlling polyvariance and global termination.

The problem of polyvariant specialisation also occurs in offline specialisation; the unfolding of a program point marked as static 
by a binding time analysis must be accompanied by an assurance that only a finite number of static instances will arise
during specialisation, and thus a finite number of versions of the program point will be generated.  This is known as bounded static variation 
\cite{Jones-Gomard-Sestoft}.

The algorithm presented here is based on a framework for partial evaluation of logic programs originally formulated in
\cite{DBLP:conf/slp/BenkerimiL90} and refined in \cite{gallagher:pepm93}, based on the framework presented by Lloyd and Shepherdson \cite{Lloyd-Shepherdson-91}. Some of the definitions in Section \ref{sec:prelim}
 are from these works, adapted for constrained Horn clauses. 

The concept of property-based abstraction has been widely used in software model checking and was first introduced
by Ball \emph{et al}. \cite{DBLP:conf/tacas/BallPR01} where it is called the Cartesian abstraction. It is used in a form similar to
that shown in this paper in the HSF tool \cite{GrebenshchikovLPR12}, though that work does not use the negations
of the properties as we do (following \cite{DBLP:conf/tacas/BallPR01}). Although we do not present an abstract interpretation \cite{Cousot-Cousot},  the domain of properties based on a finite set of constraints forms a lattice and the process of abstracting a concrete property is an example of a Galois connection. Being a finite domain, it has both
advantages and disadvantages compared to other domains that could be used to control polyvariance.
Property-based abstraction places a bound on the number of realisable
versions, whereas with an infinite-height abstract domain, with global termination ensured by widening,
the dependence of global termination on local unfolding would be loosened and an unbounded number of versions could
be produced. A general presentation of the
relation between fixed height versus infinite height domains can be found in \cite{Cousot-Cousot-92}, while an
attempt to implement global control using an infinite height lattice is shown in \cite{DBLP:journals/lisp/GallagherP01}.

Further research is needed on the automatic generation of properties. In the applications discussed in Section \ref{applications}, properties were generated automatically using various heuristics.  

Property-based abstractions are indirectly related to trace-based abstractions, which have been used in partial evaluation and supercompilation to control polyvariance, e.g. \cite{Turchin-88,Leuschel-Martens-96,Gallagher-Lafave-Dagstuhl,DBLP:conf/lopstr/AngelisFPP12}. Properties determine traces and vice versa;  a property constrains the feasible program traces; whereas a trace implicitly defines properties which permit the trace (a principle explicitly used by de Angelis \emph{et al.} \cite{DBLP:conf/lopstr/AngelisFPP12}). We consider property-based abstractions
to have some practical and conceptual advantages, opening up the use of satisfiability solvers to compute the abstraction. 
Further evaluation and investigation on the choice of properties are needed.  

The usefulness of specialisation as a component in program verification tools has been established in many works, including
\cite{deWaal-Gallagher-CADE12,Leuschel-Massart-LOPSTR99,DBLP:journals/scp/AngelisFPP14,DBLP:journals/tplp/KafleGGS18} to name only a few. Fioravanti \emph{et al.} investigated the trade-offs of polyvariance with efficiency and precision when using 
specialisation as a verification tool \cite{DBLP:journals/fuin/FioravantiPPS13}. The use of constrained Horn clauses as a semantic representation formalism for verification of a wide range of
languages and systems is now well established \cite{GrebenshchikovLPR12,DBLP:conf/birthday/BjornerGMR15}.
Here, we emphasise the
role of polyvariant specialisation in generating multiple versions of program points in a controlled way, when such versions lead to different control flow.
This in turn implicitly allows disjunctive properties to be handled.

\paragraph{Acknowledgements.} The author acknowledges very useful discussions with Bishoksan Kafle, Pierre Ganty,
Samir Genaim and Jes\'{u}s Dom\'{e}nech, contributing to the ideas presented here, as well as improvements suggested by the reviewers.

\bibliographystyle{eptcs}

\end{document}